\documentclass[12pt]{article}
\usepackage{amsmath}
\usepackage{epsfig}

\begin{document}

\author{M. O. Hase, S. R. Salinas, T. Tom\'e, and M. J. de Oliveira \\
Instituto de F\'{\i}sica, Universidade de S\~ao Paulo, \\
Caixa Postal 66318 \\
05315-970, S\~ao Paulo, SP, Brazil}
\title{The fluctuation-dissipation theorem and the linear Glauber model}
\date{\today}
\maketitle

\begin{abstract}
We obtain exact expressions for the two-time autocorrelation and response
functions of the $d$-dimensional linear Glauber model. Although this linear
model does not obey detailed balance in dimensions $d\geq 2$, we show that
the usual form of the fluctuation-dissipation ratio still holds in the
stationary regime. In the transient regime, we show the occurence of aging,
with a special limit of the fluctuation-dissipation ratio, $X_{\infty }=1/2$%
, for a quench at the critical point.


emails: mhase@if.usp.br
\end{abstract}

\section{Introduction}

It is widely recognized that the ferromagnetic Ising chain with
first-neighbor interactions and Glauber dynamics is one of the simplest,
exactly soluble, stochastic dynamical systems \cite{glauber63}. At finite
temperatures, in the stationary regime, the two-time spin autocorrelation, $%
C\left( t,t^{\prime }\right) $, and the associated response function, $%
R\left( t,t^{\prime }\right) $, of this Glauber chain are
time-translationally invariant, and duly related by the usual expression of
the fluctuation-dissipation theorem \cite{godreche00}. There is also an
aging regime, with violation of the usual form of the
fluctuation-dissipation theorem. At the critical point, at zero temperature,
for large values of the observation time $t$, the fluctuation-dissipation
ratio of the Glauber chain assumes the non trivial limoting value $X_{\infty
}=1/2$.

We now revisit a linearized version of the Glauber model, proposed by one of
us a few years ago \cite{oliveira03}. As in the original Glauber model on a $%
d$-dimensional hypercubic lattice, we still consider a one-flip stochastic
process. Each site $r=1,...,N$ of the lattice is associated with a spin
variable $\sigma _{r}=\pm 1$. However, the time evolution is now governed by
a linear spin-flip ratio, 
\begin{equation}
w_{r}(\sigma )=\frac{\alpha }{2}\left[ 1-\frac{\lambda }{2d}\sigma
_{r}\sum_{\delta }\sigma _{r+\delta }\right] ,  \label{rate}
\end{equation}
where $\lambda \in \lbrack 0,1]$ is a parameter, the sum is over the $2d$
nearest neighbors of site $r$, the time scale is set by the parameter $%
\alpha $, and $\sigma $ stands for a configuration of spin variables, $%
\sigma =\{\sigma _{r}\}$. The evolution of the probability $P(\sigma ,t)$ of
the spin configuration $\sigma $ at time $t$ is given by the master equation 
\begin{equation}
\frac{d}{dt}P(\sigma ,t)=\sum_{r\in \Lambda }\left[ w_{r}(\sigma
^{r})P(\sigma ^{r},t)-w_{r}(\sigma )P(\sigma ,t)\right] ,  \label{master}
\end{equation}
where $\sigma ^{r}$ is defined as the configuration $\sigma $ with $\sigma
_{r}$ replaced by $-\sigma _{r}$. From these equations, it is
straightforward to calculate exact analytical expression for the site
magnetization, $m_{r}(t)=\langle \sigma _{r}(t)\rangle $, and the pair
correlation function, $q_{r,r^{\prime }}(t)=\langle \sigma _{r}(t)\sigma
_{r^{\prime }}(t)\rangle $. In contrast to the original Glauber model,
defined by a non linear transition rate, with exact solutions restricted to $%
d=1$, we can now write expressions for $m_{r}(t)$ and $q_{r,r^{\prime }}(t)$
in all dimensions $d$. For $0\leq \lambda <1$, the linearized model displays
a disordered (paramagnetic) phase, with exponentially decaying pair
correlations; at the critical point $\lambda =1$, correlations decay
algebraically \cite{oliveira03}.

It is important to distinguish between this linear Glauber model and the
so-called Glauber dynamics, associated with a nonlinear transition rate, and
which is used to simulate the Ising model. The linear Glauber model may be
regarded as a voter model with noise \cite{tome89}. It displays just one
phase, for all dimensions, as long as noise is finite. In the absence of
noise ($\lambda =1$) it becomes critical. In one dimension there is a close
identification between these two versions of the Glauber model. For $d\geq 2$%
, however, in contrast to the original model, the analytically solvable
linear Glauber model is microscopically irreversible (in other words,
although having a stationary state, it does not obey detailed balance, and
cannot be associated with a Hamiltonian). The fluctuation-dissipation
theorem is usually conceived for systems that do obey detailed balance \cite%
{callen51,kubo57}. It is then reasonable to ask some questions, including
the validity of the fluctuation-dissipation theorem and the presence of an
aging regime, about the dynamical behavior of systems that do not obey
detailed balance. One of the purposes of this article is to carry out a
thorough analytical investigation of a particular system, as the non linear
Glauber model in $d\geq 2$ dimensions, which belongs to the large class of
microscopically irreversible models \cite%
{oliveira03,tome89,tome91,oliveira92,oliveira93}.

The dynamical calculations of interest in this investigation are performed
in the presence of a (small) perturbation. In the treatment of stochastic
models it is natural to introduce the modified one-spin-flip rate, 
\begin{equation}
w_{r}(\sigma )=w_{r}^{o}(\sigma )e^{-h_{r}\sigma _{r}},  \label{1}
\end{equation}
where $w_{r}^{o}\left( \sigma \right) $ is the unperturbed flipping rate
associated with the $r$th spin, and $h_{r}$ is the time-dependent
disturbance coupled to the dynamic variable $\sigma _{r}$. Alternatively, as
the calculations are restricted to small perturbations, we can write 
\begin{equation}
w_{r}(\sigma )=w_{r}^{o}(\sigma )(1-h_{r}\sigma _{r}).  \label{2}
\end{equation}
If the model is microscopically reversible, that is, if the unperturbed
transition rate $w_{r}^{o}$ obeys detailed balance, there is a model
Hamiltonian $\mathcal{H}_{o}$, and it is straightforward to show that the
perturbed transition rate $w_{r}$, given by Eq. (\ref{1}), also obeys
detailed balance. In this reversible case, the model is described by the
Hamiltonian $\mathcal{H}=\mathcal{H}_{o}-\sum_{r}H_{r}\sigma _{r}$, where $%
H_{r}=h_{r}/\beta $, and is $\beta $ proportional to the inverse of the
temperature. This form of perturbation is then suitable for reversible
models, with a disturbance $h_{r}$ proportional to the external field.
Assuming this expression for the flipping rate, the fluctuation-dissipation
theorem is given by 
\begin{equation}
R(t,t^{\prime })=\frac{\partial }{\partial t^{\prime }}\,C(t,t^{\prime }),
\label{2a}
\end{equation}
where 
\begin{equation}
R(t,t^{\prime })=\frac{1}{N}\sum_{r}\left. \frac{\delta m_{r}(t)}{\delta
h_{r}(t^{\prime })}\right| _{h\downarrow 0}  \label{R}
\end{equation}
is the response function, $m_{r}(t)$ is the average of $\sigma _{r}$ at the
observation time $t$, and 
\begin{equation}
C(t,t^{\prime })=\frac{1}{N}\sum_{r}\langle \sigma _{r}(t)\sigma
_{r}(t^{\prime })\rangle  \label{C}
\end{equation}
is the autocorrelation function of $\sigma _{r}$ between the observation
time $t$ and the waiting time $t^{\prime }$ (with $t\geq t^{\prime }$). For
systems obeying detailed balance, we have $h_{r}=\beta H_{r}$, and relation (%
\ref{2a}) reduces to the usual form of the fluctuation-dissipation theorem.
Another version of the fluctuation-dissipation theorem relates the
susceptibility, associated with a given dynamical variable $M$, such as the
total magnetization, and its variance, 
\begin{equation}
\frac{d}{dh}\langle M\rangle =\langle M^{2}\rangle -\langle M\rangle ^{2},
\label{2b}
\end{equation}
where $h$ is a static, time-independent disturbance, introduced by the
prescription of Eq. \eqref{1}.

In this work, we show that both forms of the fluctuation-dissipation
relation, \eqref{2a} and \eqref{2b}, are valid for the linear Glauber model
in the stationary regime. In the transient regime, where aging behavior
takes place, these relations are no longer obeyed. It is then appropriate to
define \cite{cugliandolo94,cugliandolo94b} a fluctuation-dissipation ratio, 
\begin{equation}
X(t,t^{\prime })=\frac{R(t,t^{\prime })}{\partial C(t,t^{\prime })/\partial
t^{\prime }}.
\end{equation}
In the linear Glauber model, for all values of the dimension $d$, we show
that $X(t,t^{\prime })\rightarrow 1$ in the limit $t^{\prime }\rightarrow
\infty $, except at the critical point, $\lambda =1$, in which case $%
X(\infty ,t^{\prime })\rightarrow 1/2$.

The layout of this paper is as follows. Some results for the linear Glauber
model, including a discussion of the lack of detailed balance for $d\geq 2$,
and calculations of the site magnetization and the spatial two-body
correlations, are reviewed is Section 2. These one-time functions play a
major role in the calculations of the two-time functions, as the
autocorrelation and the response functions, which are obtained in Section 3.
In this Section, we also consider the stationary limit and make a number of
comments on the non-stationary regime. Section 4 contains some conclusions.

\section{The linear Glauber model}

We have already mentioned that the linear Glauber model has been introduced
by one of us \cite{oliveira03} as an extension of the original Glauber model 
\cite{glauber63}, and may be regarded as a voter model with noise. This
linear Glauber model is defined by the (linear) one-spin-flip rate given by
Eq. (\ref{rate}), which should be inserted into the master equation (\ref%
{master}).

>From equations (\ref{rate}) and (\ref{master}), it is not difficult to
obtain the evolution equations for the site magnetization and the spatial
pair correlation, 
\begin{equation}
\frac{1}{\alpha }\frac{d}{dt}m_{r}(t)=-m_{r}(t)+\frac{\lambda }{2d}%
\sum_{\delta }m_{r+\delta }(t)  \label{4a}
\end{equation}
and 
\begin{equation}
\frac{1}{\alpha }\frac{d}{dt}q_{r,r^{\prime }}(t)=-2q_{r,r^{\prime }}(t)+%
\frac{\lambda }{2d}\sum_{\delta }\left[ q_{r,r^{\prime }+\delta
}(t)+q_{r^{\prime },r+\delta }(t)\right] ,  \label{4b}
\end{equation}
for $r\neq r^{\prime }$. If $r$ and $r^{\prime }$ are nearest-neighbor
sites, the right-hand side of Eq. (\ref{4b}) contains terms like $q_{r,r}(t)$
and $q_{r^{\prime },r^{\prime }}(t)$ which should be set equal to $1$. The
possibility of obtaining these exact expressions, and of performing the
exact calculations that are going to be reported in this article, is one of
the most relevant features of the linear Glauber model. As shown by Oliveira 
\cite{oliveira03}, for $0\leq \lambda <1$ this model displays a disordered
(paramagnetic) phase with exponentially decaying correlations. For $\lambda
\rightarrow 1$, it becomes critical, with algebraically decaying
correlations at $\lambda =1$.

\begin{figure}[tbp]
\begin{center}
\epsfig{file=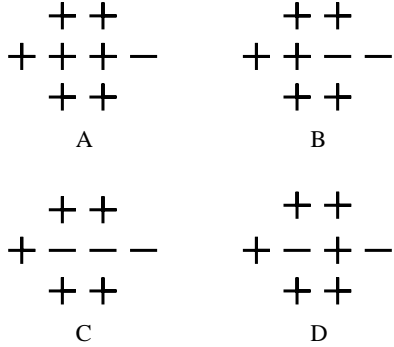,scale = 0.6}
\end{center}
\caption{A possible irreversible sequence.}
\label{seq}
\end{figure}

The linear Glauber model in one dimension has a reversible dynamics. In one
dimension, the probability of occurrence of any sequence of states and the
probability of the associated reverse sequence of states are the same. In
dimensions larger than one, this is no longer valid. Consider, for instance,
the four states shown in Fig. \ref{seq}, on a square lattice. Suppose that
the system follows the sequence of states $A$, $B$, $C$, $D$, and returns to
the initial state $A$. If the interval $\Delta t$ between two successive
states is small, then, according to the spin-flip rate given by Eq. (\ref{1}%
), the probability of occurrence of the sequence A$\rightarrow $B$%
\rightarrow $C$\rightarrow $D$\rightarrow $A is given by 
\begin{eqnarray}
P(A\rightarrow B\rightarrow C\rightarrow D\rightarrow A)
&=&P(A|D)P(D|C)P(C|B)P(B|A)P(A)  \notag \\
&=&\frac{1}{16}\left( 1-\frac{\lambda }{2}\right) ^{2}\left( 1+\lambda
\right) (\alpha \Delta t)^{4}P(A).  \notag  \label{abcda}
\end{eqnarray}
On the other hand, the reversed sequence, A$\rightarrow $D$\rightarrow $C$%
\rightarrow $B$\rightarrow $A, has the probability 
\begin{eqnarray}
P(A\rightarrow D\rightarrow C\rightarrow B\rightarrow A)
&=&P(A|B)P(B|C)P(C|D)P(D|A)P(A)  \notag \\
&=&\frac{1}{16}\left( 1+\frac{\lambda }{2}\right) ^{2}\left( 1-\lambda
\right) (\alpha \Delta t)^{4}P(A).  \notag  \label{adcba}
\end{eqnarray}
These two probabilities are distinct, so that the linear Glauber model on a
square lattice is indeed irreversible, the only exception being the trivial
case $\lambda =0$. A generalization of this result to larger dimensions is
easily found by filling the sites created by the introduction of more
dimensions with ``$+$ spins''. Hence, the detailed balance cannot be valid,
and the stationary state is \textit{a priori} not known. The connection of
the transition rates with a Gibbs measure, as it has been possible in the
one-dimensional case, is now forbidden.

\subsection{Site magnetization}

We now introduce the Fourier transform of $m_{r}(t)$, 
\begin{equation}
\tilde{m}_{k}(t)=\sum_{r}m_{r}(t)\,e^{-irk},  \label{8}
\end{equation}
and the Laplace transform of $\tilde{m}_{k}(t)$, 
\begin{equation}
\hat{m}_{k}(s)=\int_{0}^{\infty }dt\,e^{-st}\,\tilde{m}_{k}(t).  \label{9}
\end{equation}
Using these transforms, the differential equation \eqref{4a} is reduced to
the algebraic form 
\begin{equation}
\hat{m}_{k}(s)=\frac{\hat{m}_{k}^{0}}{s+\alpha f(k)},  \label{10}
\end{equation}
where 
\begin{equation}
f(k)=1-\frac{\lambda }{d}\sum_{j=1}^{d}\cos k_{j},  \label{11}
\end{equation}
and $\hat{m}_{k}^{0}$ is the Fourier-Laplace transform of $m_{r}(0)$.

The inverse Laplace transformation leads to 
\begin{equation}
\tilde{m}_{k}(t)=e^{-\alpha f(k)t}\tilde{m}_{k}(0).  \label{12}
\end{equation}
Inverse Fourier transforming, we then have 
\begin{equation}
m_{r}(t)=\sum_{r^{\prime }}\Gamma _{r-r^{\prime }}(t)m_{r^{\prime }}(0),
\label{14}
\end{equation}
where 
\begin{equation}
\Gamma _{r}(t)=\int e^{irk-\alpha f(k)t}\frac{d^{d}k}{(2\pi )^{d}}.
\label{15}
\end{equation}
We can also obtain the site magnetization with reference to an initial time $%
t=t^{\prime }$ instead of $t=0$. In this case, we just write 
\begin{equation}
m_{r}(t)=\sum_{r^{\prime }}\Gamma _{r-r^{\prime }}(t-t^{\prime
})m_{r^{\prime }}(t^{\prime }),  \label{16}
\end{equation}
where $m_{r}(t^{\prime })$ is the site magnetization at time $t^{\prime }$.

\subsection{Pair correlation}

If we look for translationally invariant solutions of Eq. \eqref{4b}, the
spatial correlation between sites $r$ and $r^{\prime }$ will be a function
of distance, $\langle \sigma _{r}(t)\sigma _{r^{\prime }}(t)\rangle
=q_{r-r^{\prime }}(t)=q_{r^{\prime }-r}(t)$. We then write Eq. \eqref{4b} as 
\begin{equation}
\frac{1}{\alpha }\frac{d}{dt}q_{r}(t)=-2q_{r}(t)+\frac{\lambda }{d}%
\sum_{\delta }q_{r+\delta }(t),  \label{6x}
\end{equation}
for $r\neq 0$, with $q_{0}(t)=1$ whenever it appears on the right-hand side.

Using a method introduced by Oliveira \cite{oliveira03}, let us write an
equation for $r=0$, 
\begin{equation}
\frac{1}{\alpha }\frac{d}{dt}q_{0}(t)=-2q_{0}(t)+\frac{\lambda }{d}%
\sum_{\delta }q_{\delta }(t)+b(t),  \label{10x}
\end{equation}
where $b(t)$ is chosen to ensure that $q_{0}(t)=1$. Actually, $b(t)$ is
defined by 
\begin{equation}
b(t)=2-\frac{\lambda }{d}\sum_{\delta }q_{\delta }(t).  \label{10y}
\end{equation}
Consequently, Eqs. \eqref{6x} and \eqref{10x} can be written as 
\begin{equation}
\frac{1}{\alpha }\frac{d}{dt}q_{r}(t)=-2q_{r}(t)+\frac{\lambda }{d}%
\sum_{\delta }q_{r+\delta }(t)+b(t)\delta _{r,0},  \label{10z}
\end{equation}
for all values of $r$.

Eq. \eqref{10z} can be examined from two points of view. On the one hand, it
is a first-order differential equations in time. On the other hand, the
summation over nearest neighbors resembles a discrete lattice Laplacian, and
from this point of view it is a discrete second-order difference equation
that can be solved by the use of a Green function. We then introduce the
Laplace transform of $q_{r}(t)$, 
\begin{equation}
\hat{q}_{r}(s)=\int_{0}^{\infty }dt\,e^{-st}\,q_{r}(t)\,,  \label{8x}
\end{equation}
so that Eq. \eqref{10z} may be written as 
\begin{equation}
\frac{1}{\alpha }\left[ -m_{0}^{2}+s\hat{q}_{r}(s)\right] =-2\hat{q}_{r}(s)+%
\frac{\lambda }{d}\sum_{\delta }\hat{q}_{r+\delta }(s)+\hat{b}(s)\delta
_{r,0},  \label{11x}
\end{equation}
where $\hat{b}(s)$ is chosen such that $q_{0}(t)=1$, which is equivalent to
take $\hat{q}_{0}(s)=1/s$. For a random initial condition, which corresponds
to a quench from infinite temperature, it is appropriate to take $%
q_{r}(0)=m_{0}^{2}\left( 1-\delta _{r,0}\right) +\delta _{r,0}$.

To solve equation \eqref{11x}, we introduce the lattice Green function
associated with a $d$-dimensional hypercubic lattice, 
\begin{equation}
\hat{G}_{r}(s)=\alpha \int \frac{e^{ikr}}{s+2\alpha f(k)}\frac{d^{d}k}{(2\pi
)^{d}},  \label{13x}
\end{equation}
where the integration is over the first Brillouin zone, and 
\begin{equation}
f(k)=1-\frac{\lambda }{d}\sum_{j=1}^{d}\cos k_{j}.  \label{13a}
\end{equation}
This Green function satisfies the equation 
\begin{equation}
\frac{1}{\alpha }s\hat{G}_{r}(s)=-2\hat{G}_{r}(s)+\frac{\lambda }{d}%
\sum_{\delta }\hat{G}_{r+\delta }(s)+\delta _{r,0}.  \label{14x}
\end{equation}

In terms of this lattice Green function, the solution of Eq. \eqref{11} can
be written as 
\begin{equation}
\hat{q}_{r}(s)=\frac{m_{0}^{2}}{s+2\alpha (1-\lambda )}+\hat{b}(s)\hat{G}%
_{r}(s)  \label{15x}
\end{equation}
where $\hat{b}(s)$ should be chosen so that $\hat{q}_{0}(s)=1/s$. This leads
to 
\begin{equation}
\hat{b}(s)=\frac{1}{\hat{G}_{0}(s)}\left[ \frac{1}{s}-\frac{m_{0}^{2}}{%
s+2\alpha (1-\lambda )}\right] ,  \label{16x}
\end{equation}
from which it follows \cite{oliveira03} the solution 
\begin{equation}
\hat{q}_{r}(s)=\frac{m_{0}^{2}}{s+2\alpha (1-\lambda )}\left[ 1-\frac{\hat{G}%
_{r}(s)}{\hat{G}_{0}(s)}\right] +\frac{\hat{G}_{r}(s)}{s\hat{G}_{0}(s)}.
\label{17a}
\end{equation}

For a completely random initial condition, $m_{0}=0$, we have $m(t)=0$, 
\begin{equation}
\hat{b}(s)=\frac{1}{s\hat{G}_{0}(s)},
\end{equation}
and 
\begin{equation}
\hat{q}_{r}(s)=\frac{\hat{G}_{r}(s)}{s\hat{G}_{0}(s)}.  \label{17b}
\end{equation}

\section{Two - time response and autocorrelation functions}

The calculation of the two-time response function, 
\begin{equation}
R(t,t^{\prime })=\frac{1}{N}\sum_{r\in \Lambda }\left. \frac{\delta m_{r}(t)%
}{\delta h_{r}(t^{\prime })}\right| _{h\downarrow 0},
\end{equation}
requires the application of a small perturbation, which is introduced
according to the prescription \eqref{2}, and from which one measures the
response of the system. As pointed out in the Introduction, we assume a
perturbed spin-flip rate, given by 
\begin{equation}
w_{r}(\sigma )=\frac{\alpha }{2}\left[ 1-\frac{\lambda }{2d}\sigma
_{r}\sum_{\delta }\sigma _{r+\delta }\right] \left[ 1-h_{r}(t)\sigma _{r}%
\right] ,  \label{25}
\end{equation}
where $h_{r}(t)$ is a time-dependent disturbance coupled to the dynamic
variable $\sigma _{r}(t)$. We then write the equation of motion for the site
magnetization, 
\begin{equation}
\frac{1}{\alpha }\frac{d}{dt}m_{r}(t)=-m_{r}(t)+\frac{\lambda }{2d}%
\sum_{\delta }m_{r+\delta }(t)+h_{r}(t)\left[ 1-\frac{\lambda }{2d}%
\sum_{\delta }q_{\delta }(t)\right] ,  \label{26}
\end{equation}
which can also be written as 
\begin{equation}
\frac{1}{\alpha }\frac{d}{dt}m_{r}(t)=-m_{r}(t)+\frac{\lambda }{2d}%
\sum_{\delta }m_{r+\delta }(t)+\frac{1}{2}h_{r}(t)b(t),  \label{26a}
\end{equation}
where $b(t)$ is given by Eq. \eqref{10y}.

Using the same procedures adopted to solve Eq. \eqref{4a}, and taking into
account that the last term in Eq. \eqref{26a} is a known function of time,
we have 
\begin{equation}
m_{r}(t)=\sum_{r^{\prime }}\Gamma _{r-r^{\prime }}(t)m_{r^{\prime }}(0)+%
\frac{\alpha }{2}\sum_{r^{\prime }}\int_{0}^{t}\Gamma _{r-r^{\prime
}}(t-t^{\prime })h_{r^{\prime }}(t^{\prime })b(t^{\prime })dt^{\prime }.
\label{27}
\end{equation}
>From this expression, we calculate 
\begin{equation}
\frac{\delta m_{r}(t)}{\delta h_{r^{\prime }}(t^{\prime })}=\frac{\alpha }{2}%
\Gamma _{r-r^{\prime }}(t-t^{\prime })b(t^{\prime }),  \label{28}
\end{equation}
which leads to the response function, 
\begin{equation}
R(t,t^{\prime })=\frac{\alpha }{2}\Gamma _{0}(t-t^{\prime })b(t^{\prime }).
\label{29}
\end{equation}

The correlation $\langle \sigma _{r}(t)\sigma _{r}(t^{\prime })\rangle $ of
a spin at a given site $r$, at time $t^{\prime }$, with the same spin at a
later time $t$ ($t\geq t^{\prime }$) is formally written as 
\begin{equation}
\langle \sigma _{r}(t)\sigma _{r}(t^{\prime })\rangle =\sum_{\sigma
}\sum_{\sigma ^{\prime }}\sigma _{r}\left( t\right) P(\sigma ,t|\sigma
^{\prime },t^{\prime })\sigma _{r}^{\prime }\left( t^{\prime }\right)
P(\sigma ^{\prime },t^{\prime }),  \label{32}
\end{equation}
where $P(\sigma ,t|\sigma ^{\prime },t^{\prime })$ is the conditional
probability of finding the configuration $\sigma $ at time $t$ given the
configuration $\sigma ^{\prime }$ at an earlier time $t^{\prime }$. Noting
that the site magnetization, $m_{r}(t)=\left\langle \sigma
_{r}(t)\right\rangle $, with the initial condition $m_{r}(t^{\prime
})=\sigma _{r}^{\prime }(t^{\prime })$, may be written as 
\begin{equation}
\sum_{\sigma }\sigma _{r}P(\sigma ,t|\sigma ^{\prime },t^{\prime })=m_{r}(t),
\label{33}
\end{equation}
and using Eq. \eqref{16}, we have 
\begin{equation}
m_{r}(t)=\sum_{r^{\prime }}\Gamma _{r-r^{\prime }}(t-t^{\prime })\sigma
_{r^{\prime }}(t^{\prime }),  \label{35}
\end{equation}
which can be inserted into Eq. \eqref{32} to give 
\begin{equation}
\langle \sigma _{r}(t)\sigma _{r}(t^{\prime })\rangle =\sum_{\sigma ^{\prime
}}\sum_{r^{\prime }}\Gamma _{r-r^{\prime }}(t-t^{\prime })\sigma _{r^{\prime
}}^{\prime }(t^{\prime })\sigma _{r}^{\prime }P(\sigma ^{\prime },t^{\prime
}),  \label{36}
\end{equation}
which finally leads to 
\begin{equation}
C(t,t^{\prime })=\sum_{r}\Gamma _{r}(t-t^{\prime })q_{r}(t^{\prime }).
\label{37}
\end{equation}

\subsection{Stationary regime}

In the stationary regime the waiting time $t^{\prime }$ and the observation
time $t$ grow without limits, but the difference $t-t^{\prime }$ is fixed.
To be more precise, $t^{\prime }\rightarrow \infty $, with $t\geq t^{\prime
} $ and $\tau =t-t^{\prime }$ fixed.

>From the Laplace transform final value theorem, we have 
\begin{equation}
q_{r}(\infty )=\lim_{t\rightarrow \infty }q_{r}(t)=\lim_{s\rightarrow 0}s%
\hat{q}_{r}(s)=\frac{\hat{G}_{r}(0)}{\hat{G}_{0}(0)},  \label{41}
\end{equation}
so that 
\begin{equation}
C(t,t^{\prime })=C(\tau )=\sum_{r}\Gamma _{r}(\tau )\frac{\hat{G}_{r}(0)}{%
\hat{G}_{0}(0)},  \label{42}
\end{equation}
which can be written as 
\begin{equation}
C(\tau )=\frac{1}{\hat{G}_{0}(0)}\int \frac{1}{2f(k)}e^{-\alpha f(k)\tau }%
\frac{d^{d}k}{(2\pi )^{d}}.  \label{43}
\end{equation}

On the other hand, taking into account that 
\begin{equation}
b(\infty )=\lim_{t\rightarrow \infty }b(t)=\lim_{s\rightarrow 0}s\hat{b}(s)=%
\frac{1}{\hat{G}_{0}(0)},
\end{equation}
the response function \eqref{29} can be written as 
\begin{equation}
R(t,t^{\prime })=R(\tau )=\frac{\alpha }{2}\Gamma _{0}(\tau )\frac{1}{\hat{G}%
_{0}(0)}.  \label{44}
\end{equation}
Using the definition of $\Gamma _{r}(t)$, given by \eqref{15}, we have 
\begin{equation}
R(\tau )=\frac{\alpha }{2\hat{G}_{0}(0)}\int e^{-\alpha f(k)\tau }\frac{%
d^{d}k}{(2\pi )^{d}}.  \label{45}
\end{equation}

Both quantities, $C(\tau )$ and $R(\tau )$, are time-translationally
invariant (functions of $\tau $ only), as it should be anticipated in a
stationary regime. Moreover, the fluctuation-dissipation theorem is
trivially satisfied, with $R(\tau )=-dC(\tau )/d\tau $. For large time
differences, and $\lambda \neq 1$, it is easy to see that both the
autocorrelation and the response functions decay exponentially, according to 
$\exp \left[ -\alpha \left( 1-\lambda \right) \tau \right] $, with the
equilibration time $\tau _{eq}=1/[\alpha \left( 1-\lambda \right) ]$.

\subsection{Global fluctuation-dissipation relation}

The fluctuation-dissipation theorem can also be written in terms of global
variables, as the magnetization and the corresponding answer to a static
perturbation. Let $M$ denote the magnetization, 
\begin{equation}
M(t)=\frac{1}{N}\sum_{r}\langle \sigma _{r}(t)\rangle ,
\end{equation}
and let us consider a static homogeneous disturbance $h$, defined by Eq. %
\eqref{1}. Then 
\begin{equation}
\frac{dM(t)}{dh}=\chi (t),  \label{61}
\end{equation}
where $\chi (t)$ is a variance, 
\begin{equation}
\chi (t)=\frac{1}{N}\sum_{r}\sum_{r^{\prime }}\left[ \langle \sigma
_{r}(t)\sigma _{r^{\prime }}(t)\rangle -\langle \sigma _{r}(t)\rangle
\langle \sigma _{r^{\prime }}(t)\rangle \right] .
\end{equation}

Due to the translational invariance of the lattice, $M(t)=\langle \sigma
_{0}(t)\rangle $. This quantity is the solution of Eq. \eqref{26a}, which is
now given by 
\begin{equation}
\frac{1}{\alpha }\frac{d}{dt}M(t)=-(1-\lambda )M(t)+\frac{1}{2}hb(t),
\end{equation}
with the stationary solution 
\begin{equation}
M=h\frac{b(\infty )}{2(1-\lambda )}=h\frac{1}{2(1-\lambda )\hat{G}_{0}(0)}.
\label{53}
\end{equation}

For $\lambda <1$, the magnetization vanishes as $h\rightarrow 0$, so that
the variance is given by 
\begin{equation}
\chi (t)=\sum_{r}q_{r}(t).  \label{56}
\end{equation}
>From Eq. \eqref{10z} it follows that 
\begin{equation}
\frac{1}{\alpha }\frac{d}{dt}\chi (t)=-2(1-\lambda )\chi (t)+b(t),
\label{57}
\end{equation}
with the stationary solution 
\begin{equation}
\chi =\frac{b(\infty )}{2(1-\lambda )}=\frac{1}{2(1-\lambda )\hat{G}_{0}(0)}.
\label{58}
\end{equation}
>From equations \eqref{53} and \eqref{58}, it is seen that relation %
\eqref{61} is clearly satisfied in the stationary regime.

\subsection{Aging regime}

The temporal behavior of the autocorrelation and the response functions, as
calculated in this section, already suggests the existence of an aging
regime. Equations (\ref{29}) and (\ref{37}), for $R(t,t^{\prime })$ and $%
C(t,t^{\prime })$, are valid for all values of $t^{\prime }$ and $t$, with $%
t-t^{\prime }=\tau \geq 0$. The dependence of these functions on both $t$
and $t^{\prime }$, and not on $\tau $ only, leads to the existence of aging.
In this regime, the role of the spatial correlations is still crucial, since
they are responsible for the realization of the aging scenario. In the
stationary regime, it should be noted that $q_{r}$ becomes a
time-independent quantity in the $t^{\prime }\rightarrow \infty $ limit
only, and this is the reason of the dependence of the two-time functions on $%
\tau $ only.

In the transient regime, we do not expect the validity of the usual form of
the fluctuation-dissipation relation given by Eq. \eqref{2a}. It has been
convenient \cite{cugliandolo94,cugliandolo94b,godreche00,godreche00b,hase05}
to characterize the distance to the stationary regime by the
fluctuation-dissipation ratio, 
\begin{equation}
X(t,t^{\prime })=\frac{R(t,t^{\prime })}{\partial C(t,t^{\prime })/\partial
t^{\prime }}.
\end{equation}
A particular interesting quantity is the limit 
\begin{equation}
X_{\infty }=\lim_{t^{\prime }\rightarrow \infty }X(\infty ,t^{\prime }),
\end{equation}
where 
\begin{equation}
X(\infty ,t^{\prime })=\lim_{t\rightarrow \infty }X(t,t^{\prime }).
\end{equation}

>From the results of this Section, it is easy to write 
\begin{equation}
X(\infty ,t^{\prime })=\frac{b(t^{\prime })}{2b(t^{\prime })-2(1-\lambda
)\chi (t^{\prime })},
\end{equation}
where $\chi (t^{\prime })$ is the sum given by \eqref{56}. At the critical
point, $\lambda =1$, it follows that $X(\infty ,t^{\prime })=1/2$ for any
time $t^{\prime }$, so that $X_{\infty }=1/2$. In the disordered phase, $%
\lambda \neq 1$, Eqs. \eqref{57} and \eqref{58} may be used to conclude that 
$X_{\infty }=1$.

\section{Conclusions}

We have reported a number of exact calculations for the dynamical behavior
of a $d$-dimensional linearized version of the stochastic Glauber model. In
one dimension, both the linear and the original model are essentially
equivalent. For $d\geq 2$, however, the rates of transition of the linear
Glauber model do not obey the conditions of detailed balance. This linear
model can be regarded as a voter model with noise; it displays just one
stable phase as long as noise is finite, and becomes critical in the absence
of noise ($\lambda =1$). Since the dynamical properties are usually
conceived for systems that do obey detailed balance, we decided that it was
appropriate to carry out some explicit calculations for a microscopically
irreversible model.

We have obtained expressions for the two-time autocorrelation, $C\left(
t,t^{\prime }\right) $, and response functions, $R\left( t,t^{\prime
}\right) $, which depend on both observation time $t$ and waiting time $%
t^{\prime }\leq t$. In the stationary, infinite time limit, $t^{\prime
}\rightarrow \infty $, in the presence of noise ($0<\lambda <1$), the
spatial correlations are independent of time, and the two-time functions
become translationally invariant (depending on the difference $\tau
=t-t^{\prime }$ only). The usual form of the fluctuation-dissipation
theorem, $X(t,t^{\prime })=R(t,t^{\prime })/\partial C(t,t^{\prime
})/\partial t^{\prime }=1$, is trivially observed in this regime.

In the scaling regime ($t\rightarrow \infty $), for $0\leq \lambda <1$, we
obtain a non trivial fluctuation-dissipation ratio, $X(\infty ,t^{\prime
})\neq 1$. At the critical point, $\lambda =1$, we have $X(\infty ,t^{\prime
})=1/2$, which further indicates that the dynamical behavior of the $d$%
-dimensional microscopically irreversible linear Glauber model is very
similar to the behavior of its one-dimensional reversible
counterpart.\bigskip

\textbf{Acknowledgments}\bigskip

The authors acknowledge the financial support of the Brazilian agencies
CAPES, CNPq, and FAPESP.

\end{document}